%% file: lens.tex
\documentstyle[11pt,aaspp4,epsf]{article}
\def\vol#1  {{{#1}{\rm,}\ }}
\def\aa{{A\&A}, }     %Astronomy & Astrophysics%
\def\aasup{{A\&AS}, } %A & A Supplements%
\def\araa{{ARA\&A}, }
\def\aj{{AJ}, }  %Astronomical Journal%
\def\apj{{ApJ}, } %Astrophysical Journal%
\def\apjs{{ApJS}, } %Astrophysical Journal Supplements%
\def\pasp{{PASP}, }  %Publications of the Astronomical%
\def\mnras{{MNRAS}, } %Monthly Notices of the Royal%
\begin{document}

\title{ A Keck Survey of Gravitational Lens Systems  :  \\
        I. Spectroscopy of SBS 0909+532, HST 1411+5211, and CLASS B2319+051}

\author{L.M.\ Lubin\altaffilmark{1}}
\affil{Palomar Observatory, 105-24, California Institute of Technology, Pasadena, CA 91125}
\affil{lml@astro.caltech.edu}

\author{C.D.\ Fassnacht}
\affil{NRAO, P.O. Box 0, Socorro, NM, 87801}
\affil{cfassnac@aoc.nrao.edu}

\author{A.C.S.\ Readhead}
\affil{Palomar Observatory, 105-24, California Institute of Technology, Pasadena, CA 91125}
\affil{acr@astro.caltech.edu}

\author{R.D.\ Blandford}
\affil{Theoretical Astrophysics, 130-33, California Institute of Technology, Pasadena, CA 91125}
\affil{rdb@tapir.caltech.edu}

\author{T.\ Kundi\'{c}}
\affil{Renaissance Technologies, 600 Route 25A, East Setauket, NY 11733}

\vskip 1 cm
\centerline{Accepted for publication in the {\it Astronomical Journal}}

\altaffiltext{1}{Hubble Fellow}

\vfill
\eject

\begin{abstract}

We present new results from a continuing Keck program to study
gravitational lens systems.  We have obtained redshifts for three lens
systems, SBS 0909+532, HST 1411+5211, and CLASS B2319+051. For all of
these systems, either the source or lens redshift (or both) has been
previously unidentified. Our observations provide some of these
missing redshifts. We find $(z_{\ell}, z_s) = (0.830, 1.377)$ for SBS
0909+532; $(z_{\ell}, z_s) = (0.465, 2.811)$ for HST 1411+5211,
although the source redshift is still tentative; and $(z_{{\ell}_{1}},
z_{{\ell}_{2}}) = (0.624, 0.588)$ for the two lensing galaxies in
CLASS B2319+051.  The background radio source in B2319+051 has not
been detected optically; its redshift is, therefore, still unknown.
We find that the spectral features of the central lensing galaxy in
all three systems are typical of an early-type galaxy.  The observed
image splittings in SBS 0909+532 and HST 1411+5211 imply that the
masses within the Einstein ring radii of the lensing galaxies are $1.4
\times 10^{11}$ and $2.0 \times 10^{11}~h^{-1}~M_{\odot}$,
respectively. The resulting $B$ band mass-to-light ratio for HST
1411+5211 is $41.3 \pm 1.2~h~{(M/L)}_{\odot}$, a factor of $\sim$ 5
times higher than the average early-type lensing galaxy. This large
mass-to-light is almost certainly the result of the additional mass
contribution from the cluster CL 3C295 at $z = 0.46$. For the lensing
galaxy in SBS 0909+532, we measure ${(M/L)}_{B} =
4^{+11}_{-3}~h~{(M/L)}_{\odot}$ where the large errors are the result
of significant uncertainty in the galaxy luminosity. While we cannot
measure directly the mass-to-light ratio of the lensing galaxy in
B2319+051, we estimate that ${(M/L)}_{B}$ is between
$3-7~h~{(M/L)}_{\odot}$.

\end{abstract} 


\section{Introduction}

Gravitational lensing has proven to be an invaluable astrophysical
tool for constraining the cosmological parameters $H_{0}$ (Kundi\'c et
al.\ 1997a; Schechter et al.\ 1997; Lovell et al.\ 1998; Biggs et al.\
1999; Fassnacht et al.\ 1999) and $\Lambda$ (Falco, Kochanek \&
Mu\~noz 1998; Helbig et al.\ 1999). In addition, a unique contribution
of gravitational lensing to extragalactic astronomy lies in its
capacity to measure directly the masses of the lensing objects.
Consequently, it can be used to study galaxy structure and its
evolution with redshift (e.g.\ Keeton, Kochanek \& Falco 1998). The
advent of high-spatial-resolution imaging with HST and faint-object
spectroscopy with the Keck 10-m telescopes have opened new
possibilities in the field (e.g.\ Kundi\'c et al.\ 1997b,c; Fassnacht
\& Cohen 1998).  Systems with compact configurations and faint
components can now be studied, increasing the size and completeness of
statistical samples of lenses.  Specifically, a detailed study of a
large number of gravitational lens systems can be used (1) to identify
simple lens systems for the measurement of $H_{0}$; (2) to measure the
mass-to-light of the lensing galaxies; (3) to compare the dark matter
to the stellar light distribution of the lens galaxies; and (4) to
probe the interstellar medium in the lensing galaxies.  Nearly all of
these goals depend critically on accurate redshift determinations for
the background sources and the lensing galaxies.

In light of this, we have begun a coordinated program to use the Low
Resolution Imaging Spectrograph (LRIS; Oke et al.\ 1995) on the Keck
II telescope to measure spectroscopic redshifts for all lens systems
where either the source or lens redshift is currently unavailable. We
have drawn our sources from the sample of the CfA-Arizona Space
Telescope Lens Survey of gravitational lenses (CASTLES). The CASTLES
team has compiled a list of all known confirmed or candidate
gravitational lens systems with angular separations smaller than
$10^{''}$. These systems were originally identified by a variety of
methods and by many different groups. The specific goal of CASTLES is
the construction of a complete three-band ($V$, $I$, and $H$)
photometric survey of this sample. CASTLES uses existing Hubble Space
Telescope (HST) images when available.  Otherwise, they have
supplemented the archival data with new WFPC2 and/or NICMOS imaging
(see {\tt http://cfa-www.harvard.edu/castles}).

In addition, we have pre-publication access to new gravitational lens
candidates discovered in the Cosmic Lens All-Sky Survey (CLASS).  The
CLASS survey is being conducted at radio wavelengths with the VLA and
consists of observations of $\sim$12,000 flat-spectrum radio sources
to search for gravitational lens candidates.  The first three phases
of this survey have confirmed 12 new lenses and found $\sim$10
additional candidates (Myers et al.\ 1999).  As part of the CLASS
follow-up observations, many of these lenses have been imaged in two
or three bands with HST (Jackson et al.\ 1998a,b; Koopmans et al.\
1998, 1999; Sykes et al.\ 1998; Fassnacht et al.\ 1999; Xanthopoulos
et al. 1999). At this time, eight of the 12 confirmed lenses from
CLASS are included in CASTLES.

Earlier results from the first phases of this Keck survey have already
been published (Kundi\'c et al.\ 1997b,c; Fassnacht \& Cohen 1998).
In this paper, we present spectra of three lens systems with missing
redshifts : SBS 0909+532, HST 1411+5211, and CLASS B2319+051. Unless
otherwise noted, we use $H_{0} = 100~h~{\rm km~s^{-1}~Mpc^{-1}}$,
$\Omega_{m} = 0.2$, and $\Omega_{\Lambda} = 0.0$.

\section{Targets}

Below we present some relevant information on the previous
observations of the three lens systems which are the subject of this
paper.

\subsection{SBS 0909+532}

SBS 0909+532 was first discovered as a quasar by Stepanyan et al.\
(1991) and later identified in the Hamburg-CfA Bright Quasar Survey
(Engels et al.\ 1998).  Kochanek, Falco \& Schild (1995) believed that
this quasar was a good candidate for gravitational lensing because of
its redshift ($z = 1.377$) and its bright optical magnitude ($B =
17.0$). Kochanek et al.\ (1997) first resolved this source into a
close pair which was separated by $\Delta \theta = 1\farcs{11}$ and
had a flux ratio of $R_B - R_A = 0.58$ mag.  These observations
suggested that this system was indeed a gravitational lens. Oscoz et
al.\ (1997) confirmed this hypothesis with spectra of the two
components taken at the William Herschel Telescope (WHT).  The spectra
showed that components A and B were quasars at the same redshift and
had identical spectra.  Oscoz et al.\ (1997) also detected the
\ion{Mg}{2} $\lambda \lambda 2796, 2803$ doublet in absorption at the
same redshift ($z = 0.83$) in both components. They argued that these
absorption features were associated with the photometrically
unidentified lensing galaxy. Optical and infrared HST imaging indicate
that the lensing galaxy has a large effective radius ($r_e =
1\farcs{58} \pm 0\farcs{90}$) and a correspondingly low surface
brightness. It has a total magnitude of $H = 16.75 \pm 0.74$ and a
color of $I-H = 2.28 \pm 1.01$ within an aperture of diameter
1\farcs{7} (Leh\'ar et al.\ 1999). The large uncertainties are a
result of the difficulty in subtracting the close pair of quasar
images (see Figure 1 of Leh\'ar et al.\ 1999).  Our observations
confirm that the lensing galaxy is at the same redshift as the
\ion{Mg}{2} absorbers.

\subsection{HST 1411+5211}

HST 1411+5211 is a quadruple lens that was discovered by Fischer,
Schade \& Barrientos (1998) in archival WFPC2 images taken of the
cluster CL 3C294 (CL1409+5226) with the F702W filter. The maximum
image separation is 2\farcs{28}.  The intensities of the four
components are reasonably similar; the F702W AB magnitudes correspond
to $\{{\rm A~B~C~D}\} = \{24.96~25.95~24.92~25.00\}$. The primary
lensing galaxy is clearly observed in the HST images with a total
magnitude of F702W(AB) $= 20.78 \pm 0.05$. It has the appearance of a
morphologically normal elliptical galaxy with a measured half-light
radius of $r_{1\over{2}} = 0\farcs{61} \pm 0\farcs{03}$ and an
ellipticity of $\epsilon = 0.27 \pm 0.03$. The lensing galaxy is
located only 52\farcs{0} (or $195~h^{-1}$ kpc) from the center of the
massive cluster CL 3C295 at $z = 0.46$ (Butcher \& Oemler
1978). Although this cluster was the subject of an extensive
spectroscopic survey by Dressler \& Gunn (1992), there is no measured
redshift for the lensing galaxy (identified as galaxy \#162 of Table 6
in Dressler \& Gunn 1992); however, a photometric redshift of $z =
0.598 \pm 0.11$ based on narrow-band imaging has been measured (Thimm
et al.\ 1994). Fischer et al.\ (1998) argued that this photometric
redshift was suspect. Firstly, the photometric redshift had the
largest quoted uncertainty of all the observed galaxies (over two
times larger than the average).  Secondly, Thimm et al.\ (1994)
classified this galaxy as an Scd based on their measurement of the
spectral energy distribution.  The high-angular-resolution HST imaging
clearly indicates that this galaxy is an early-type, not a late-type,
galaxy.  In this paper, we convincingly show that the photometric
redshift of Thimm et al.\ (1994) is incorrect.

\subsection{CLASS B2319+051}

B2319+051 is a doubly-imaged gravitational lens systems newly
discovered by CLASS (Marlow et al.\ 1999).  Radio images taken with
the Very Large Array (VLA) and the Multiple-Element Radio-Linked
Interferometer (MERLIN) show two compact components aligned in a N-S
orientation with a separation of 1\farcs{36} and a flux density ratio
of 5.7:1.  High-resolution radio imaging with the Very Large Baseline
Array (VLBA) resolve each component into two subcomponents with a
separation of 0\farcs{021} for A and 0\farcs{0075} for B. The
orientation and morphology of this configuration is consistent with
the lensing hypothesis. Images of this system taken with NICMOS do not
show any infrared counterparts to the radio components; however, it
does reveal two lensing galaxies (Marlow et al.\ 1999).  G1 is a
large, elliptical-like galaxy which is associated with the position of
the two radio components; hence, it is the primary lensing galaxy. G2
is an extended, irregular galaxy which shows two clear emission peaks
(G2a and G2b) and is separated from G1 by G1--G2b $=$ 3\farcs{516}
(see Figure 9 of Marlow et al.\ 1999).  This galaxy is the source of
an external shear as modeled by Marlow et al.\ (1999). The integrated
magnitudes of G1 and G2 are F160W $= 18.2$ and 19.1, respectively.

\section{Observations}

All of the observations were performed with the Low Resolution Imaging
Spectrograph (LRIS; Oke et al.\ 1995) on the Keck II telescope. For
the spectroscopic observations, we have used the instrument in
long-slit mode with the $300~{\rm grooves~mm^{-1}}$ grating which
provides a spectral resolution of $2.44~{\rm \AA~pixel^{-1}}$. The
long slit was aligned along the axis defined by the two images of the
background source for both SBS 0909+532 and CLASS B2319+051.  Note
that the latter position covers the primary lensing galaxy G1 in the
B2319+051 system but not G2.  For galaxy G2, the longslit was placed
along the axis defined by its two components, G2a and G2b (see
\S2.3). For HST 1411+5211, the long slit was aligned along the axis
defined by images A and C of the background source (see Fischer et
al.\ 1998).  Except for galaxy G2 of B2319+051 where only one exposure
was taken, two exposures of equal duration were taken for each object.
The specific details of these observations are listed in Table 1. In
addition, we have obtained $R$ images of CLASS B2319+051 using LRIS in
imaging mode. These data are the only optical imaging available on
this source. The total exposure time for these observations is 1200
sec.

In all cases, the data were reduced using standard IRAF\footnote{IRAF
is distributed by the National Optical Astronomy Observatories, which
are operated by the Association of Universities for Research in
Astronomy, Inc., under cooperative agreement with the NSF.} routines.
The bias levels were estimated from the overscan region on each chip.
For the imaging data, a flat-field was constructed from dome flats
taken in the beginning of each night.  For the spectroscopic
observations, flat-fielding and wavelength calibration were performed
using internal flat-field and arc lamp exposures which were taken
after each science exposure.  Observations of the Oke (1990)
spectrophotometric standard stars Feige 34, G138-31, BD332642, and
Feige 110 were used to remove the response function of the chip.  The
individual spectra for each object were weighted by the squares of
their signal-to-noise ratios and combined.

\section{Results}

The final spectra are shown in Figures~\ref{0909spec} --
\ref{1411source} and \ref{1411gal} -- \ref{2319spec}.  The lines used
to identify the redshifts of the lensing galaxies and the background
sources are given in Table 2.  The redshift uncertainties (see Table
3) have been estimated by taking the rms scatter in the redshifts
calculated from the individual spectral lines.  We present a more
detailed discussion of the individual systems below.

\subsection{SBS 0909+532}

The spatial projection of the spectra from the SBS 0909+532 system
shows a double peak, with the sub-peaks separated by approximately 5
pixels or 1\farcs1.  This separation matches the 1\farcs107 quasar
image separation measured by Kochanek et al.\ (1997).  The spectrum
shown in Figure~\ref{0909spec} was extracted using an aperture of 2
pixels placed in the trough between the sub-peaks of emission in order
to maximize the fractional contribution of the lensing galaxy.  The
final spectrum is still dominated by light from the background source,
a quasar at $z_s = 1.377$ with broad \ion{C}{3}] and \ion{Mg}{2}
emission lines (as seen in Oscoz et al.\ 1997).  However, it is
possible to see features from the lensing galaxy, including the
\ion{Ca}{2} H and K doublet, which establishes the lens redshift as
$z_{\ell} = 0.830$. The features identified with the lensing galaxy
are typical of an early-type galaxy.  For a non-evolving elliptical
galaxy at the lens redshift, we expect an optical--infrared color of
$I-H \sim 2$ (Poggianti 1997).  Consequently, the observed value of
$I-H = 2.28 \pm 1.01$ (Leh\'ar et al.\ 1999) provides additional
support for an early-type classification of the lensing galaxy.

\subsection{HST 1411+5211}

The spectrum of lens system HST 1411+5211 shows two distinct traces, a
bright central source which is separated by approximately 5 pixels or
1\farcs{1} from a significantly fainter one. The two traces correspond
to the lensing galaxy and the background source, respectively, as the
separation is exactly that expected from the high-angular-resolution
HST imaging (Fischer et al.\ 1998). From these spectra, we have
obtained the source and lens redshift of $z_{\ell} = 0.465$ and $z_s =
2.811$, respectively.  In the spectrum of the lensing galaxy, the
strong 4000\AA\ break, the small equivalent width Balmer absorption
lines, and the lack of [\ion{O}{2}] emission indicate that little star
formation is occurring (Figure~\ref{1411lens}). The spectral features
are consistent with the fact that this galaxy appears as a
morphologically normal elliptical.  The measured redshift proves that
the lensing galaxy is a member of the cluster CL 3C295.

The background source shows a modest emission line at an observed
wavelength of 4634\AA\ (Figure~\ref{1411source}).  This line is much
more obvious in the two-dimensional, sky-subtracted spectrum than in
this one-dimensional spectrum (see Figure~\ref{1411s2D}).  There are
only two plausible interpretations of this emission line as all other
choices would require the presence of other, stronger emission
lines. Firstly, the line could be [\ion{O}{2}] 3273\AA\ at $z_s =
0.243$.  We would then expect to see comparably strong [\ion{O}{3}]
5007\AA, 4959\AA\ at 6164\AA, 6224\AA\ or H$\beta$ 4861\AA\ at
6042\AA. None of these lines are seen in the data, although the
spectrum is much weaker at these wavelengths. This identification
would also imply that the emission is not coming from the background
source, but rather from some unrelated foreground object.  Because of
the lack of other emission lines and the exact coincidence with the
position of the background source, we believe the only reasonable
explanation for this line is Ly$\alpha$ 1215.7\AA\ at $z_s = 2.811$.
The appearance of this spectrum is similar to other known star-forming
galaxies at comparable redshifts with absorption features which
include e.g.\ \ion{Si}{2} and \ion{C}{4} (Steidel et al.\ 1996a,b).
In addition, there is a continuum break blueward of this line with a
drop amplitude (Oke \& Korycansky 1982) of $D_A = 0.25 \pm 0.05$.
This decrement is due to absorption by intervening hydrogen and is
consistent with that found in the spectra of other high-redshift
objects (e.g.\ Oke \& Korycansky 1982; Kennefick et al.\
1995). Because of the low signal-to-noise in this spectrum, we still
regard this redshift measurement as tentative.  We are planning to
re-observe this object during the next observing season.

In addition to the lens system, we have also obtained spectra of two
galaxies which happened to lie on the long slit during the
observations of the gravitational lens system.  They are identified as
galaxies \#158 and \#165 in the cluster field CL 3C295 (see Table 6 of
Dressler \& Gunn 1992).  Dressler \& Gunn (1992) list their total $r$
magnitudes as 20.13 and 22.56, respectively.  The redshift of each
galaxy was previously unknown.  Based on our spectra, we find a
redshift of $z = 0.451$ for both galaxies (Figure~\ref{1411gal}),
indicating that the galaxies are cluster members.  Each spectrum shows
the classic K star absorption features of \ion{Ca}{2} H \& K which are
typical of an early-type galaxy. In addition, they show a series of
strong Balmer absorption lines, including H$\theta$, H$\eta$,
H$\zeta$, H$\delta$, H$\gamma$, and H$\beta$, which suggest that these
galaxies are ``K+A'' (or more commonly known as ``E+A'') galaxies
(Dressler \& Gunn 1983; Gunn \& Dressler 1992; Zabludoff et al.\
1997).  These spectral features imply that these galaxies have
experienced a brief starburst within the last 1--2 Gyrs.

\subsection{CLASS B2319+051} 

We have obtained spectra of the two lensing galaxies, G1 and G2, in
B2319+051. No optical emission associated with the background radio
source has been detected; thus, the source redshift is still
unknown. The redshifts of the two lensing galaxies are
$(z_{{\ell}_{1}}, z_{{\ell}_{2}}) = (0.624, 0.588)$.  As the redshifts
indicate, G2 is not a companion galaxy to the primary lensing galaxy
G1. Rather, they are just a chance superposition along the
line-of-sight. The spectrum of G1 is consistent with its morphological
identification as an early-type galaxy in the high-angular-resolution
NICMOS image (Marlow et al.\ 1999). It has a strong 4000\AA\ break and
small equivalent width Balmer absorption lines. It does, however, show
some indication of current star formation with a modest [\ion{O}{2}]
line (equivalent width of 9\AA).  Galaxy G2 is clearly more active as
it has much stronger [\ion{O}{2}] emission (equivalent width of 22\AA)
and a less well-defined 4000\AA\ break.  In addition, the spectrum
shows a series of strong Balmer absorption features which indicates a
burst of star formation within the last 1--2 Gyrs (see e.g.\ \S
4.2). Such activity is expected as the galaxy appears morphologically
irregular with two distinct peaks in the surface brightness profile.
This appearance suggests a merger or interaction.

The composite $R$ band image of a $1' \times 1'$ field centered on
B2319+051 is shown in Figure~\ref{2319im}.  Using the object detection
and analysis software SExtractor (Bertin \& Arnouts 1996), we have
obtained the magnitude $R = 22.2 \pm 0.3$ for the primary lensing
galaxy G1 within an aperture the size of the Einstein ring radius
(0\farcs{68}). In addition, the total $R$ magnitudes of G1 and G2 are
$21.3 \pm 0.3$ and $22.0 \pm 0.3$, respectively. The errors are large
because these data were taken in non-photometric conditions with light
to moderate cirrus.  The total $R - F160W$ color of G1 is consistent
with a non-evolving elliptical at a redshift of $z = 0.624$ (Poggianti
1997).

\section{The Mass and Light}

Once the source and lens redshifts of a gravitational lens system are
known, the system can be used, in principle, for two distinct
purposes.  Firstly, it is possible to measure $H_{0}$ by combining the
angular diameter distances and a model of the lensing potential to
predict the time delays (see e.g.\ Refsdal 1964; Blandford \& Narayan
1992; Blandford \& Kundi\'c 1996). The predicted time delay is
proportional to the ratio of angular diameter distances, $D \equiv
{{D_{\ell} D_s}\over{D_{{\ell} s}}}$ (where $D_{\ell}$, $D_s$, and
$D_{{\ell} s}$ are the angular diameter distances to the lens, to the
source, and between the lens and source, respectively). As such, the
predicted time delay is also inversely proportional to $h$.  Thus, if
the background source is variable, and the time delays can be
measured, the ratio between the observed and predicted time delays
will provide a measure of $h$. Unfortunately, a time delay measurement
requires long-term radio or optical monitoring and a detection of a
relatively strong event (see e.g.\ Kundi\'c et al.\ 1997a; Schechter
et al.\ 1997; Lovell et al.\ 1998; Biggs et al.\ 1999; Fassnacht et
al.\ 1999).  Consequently, these measurements are difficult to make.

More immediately, gravitational lens systems with measured redshifts
can be used to study the properties of massive galaxies at moderate
redshift.  Specifically, the size of the image splitting provides a
direct estimate of the mass within the Einstein ring of the lens. This
mass can be expressed as :

\begin{equation}
M_E \approx 1 \times 10^{12} \left({D}\over{1~{\rm Gpc}}\right)
{\left({\Theta_E}\over{3^{''}}\right)}^{2} M_{\odot}
\end{equation}

\noindent where $\Theta_E$ is the angular radius of the Einstein
ring. For the lenses presented in this paper, we find physical
Einstein ring radii of $2.6 - 4.3~h^{-1}$ kpc and masses of $\sim 1 -
2 \times 10^{11}~h^{-1}~M_{\odot}$ (see Table 3).

The mass of the galaxy, combined with its photometric properties, can
be used to compute the mass-to-light of the lens.  For this
calculation, we need to measure the galaxy light within the same
aperture as the mass. For both SBS 0909+532 and HST 1411+5211, all of
the necessary parameters for the mass-to-light ($M/L$) calculation
have been measured.  For the remaining system B2319+051, we can only
provide a reasonable estimate.  In the calculations presented below,
all of the galaxy magnitudes are given in a Vega-based (``Johnson'')
magnitude system.  In addition, we have converted all observed
magnitudes to the rest-frame $B$ band using no-evolution $k$
corrections and rest-frame colors calculated from the spectral energy
distribution of a typical elliptical galaxy (Coleman, Wu \& Weedman
1980).  We have ignored the effects of extinction and evolution.
While the total extinction is usually modest in early-type lenses
[$E(B-V) \le 0.08$ mag; Falco et al.\ 1999], the evolutionary
correction is, as expected, an increasing function of redshift,
approaching 1 mag at redshifts of $z \sim 0.9$ (Kochanek et al.\ 1999).

\subsection{SBS 0909+532}

The properties of the lensing galaxy in SBS 0909+532 have been
measured by Leh\'ar et al.\ (1999).  They give a total magnitude of $H
= 16.75 \pm 0.74$, a color of $I-H = 2.28 \pm 1.01$ within a
1\farcs{7} diameter aperture, and an effective radius of $r_e =
1\farcs{58} \pm 0\farcs{90}$. The errors on these parameters are
extremely large because the subtraction of the close quasar pair
leaves significant residuals in the final image (see Figure 1 of
Leh\'ar et al.\ 1999). However, we can try to use these values to
estimate the light within the Einstein ring radius of 0\farcs{55}.
Adopting a de Vaucouleurs law for the galaxy surface brightness
profile, we calculate that the magnitude within the Einstein ring
radius would be $H = 18.3^{+0.9}_{-1.0}$.  If we assume that the
galaxy color is constant with radius, the $I$ magnitude corresponds to
$20.6^{+1.3}_{-1.4}$.  Converting this value to an absolute $B$
magnitude, we find $M_{B} = -20.9^{+1.4}_{-1.5} + 5~{\rm log}~h$ and
${(M/L)}_{B} = 4^{+11}_{-3}~h~{(M/L)}_{\odot}$.  Although this
measurement does not place any strong constraints on the $M/L$ of this
lensing galaxy, it is consistent with the mass-to-light ratios of
other early-type lenses at $z \sim 0.8$.  From the review of Keeton et
al.\ (1998), we would expect ${(M/L)}_{B} \approx 8 -
16~h~{(M/L)}_{\odot}$.  We note that the mass-to-light ratios of
high-redshift lensing galaxies are higher (by a factor of $\sim 1.5 -
2$) than the $M/L$ ratios of nearby elliptical galaxies within the
same physical radius (e.g.\ Lauer 1985; van der Marel 1991); however,
searches for gravitational lenses are biased toward high mass systems
since these systems have a larger cross-section for lensing.

\subsection{HST 1411+5211}

For HST 1411+5211, we have obtained the photometry of the lensing
galaxy from the processed WFPC2 image of the cluster CL 3C295 which is
given in Smail et al.\ (1997). We adopt a zero point in the F702W
bandpass of $22.38 \pm 0.02$ mag DN$^{-1}$ s$^{-1}$ (Holtzman et al.\
1995) and measure an aperture magnitude of F702W $= 21.23 \pm 0.03$
within the Einstein ring radius of 1\farcs{14}. Converting this value
to an absolute $B$ magnitude, we find $M_{B} = -18.72 \pm 0.03 +
5~{\rm log}~h$ and ${(M/L)}_{B} = 41.3 \pm
1.2~h~{(M/L)}_{\odot}$. This mass-to-light ratio is considerably
higher (by a factor of $\sim 5$) than the average lensing galaxy at $z
\sim 0.4$ (Keeton et al.\ 1998). The inflated value is the result of
cluster--assisted galaxy lensing induced by the cluster CL 3C295; this
cluster is extremely massive with a velocity dispersion of $\sigma =
1670~{\rm km~s^{-1}}$ (Dressler \& Gunn 1992).  Such an effect is also
seen in the gravitational lens system Q0957+561 where the contribution
of the $\sigma = 730~{\rm km~s^{-1}}$ cluster (Angonin-Williame,
Soucail \& Vanderriest 1994; Fischer et al.\ 1997) results in an
unusually high value of ${(M/L)}_{B} \approx 22~h$ for the central
lensing galaxy G1 (Keeton et al.\ 1998).

\subsection{CLASS B2319+051}

For B2319+051, we have calculated an aperture magnitude of $R = 22.2
\pm 0.3$ for the lensing galaxy G1 (see \S 4.3).  This magnitude
corresponds to $M_{B} = -20.4 \pm 0.3 + 5~{\rm log}~h$ or a luminosity
of $L_B = 2.3 \pm 0.6 \times 10^{10}~h^{-2}~L_{\odot}$. Because the
redshift of the background source in this system is not known, we
cannot calculate the mass-to-light ratio of the lensing galaxy.
However, using the measured luminosity and equation (1), we can
represent the $M/L$ ratio of G1 as a function of ${D_s}
\over{D_{{\ell} s}}$.  That is,

\begin{equation}
{(M/L)}_{B} \approx 2.00 \left({D_s}\over{D_{{\ell} s}}\right)
~h~{(M/L)}_{\odot}
\end{equation}

\noindent For reasonable values of the source redshift i.e.\ $z_s =
1-3$, we estimate that ${(M/L)}_{B}$ will be between $7 -
3~h~{(M/L)}_{\odot}$.  In our chosen cosmology, all other lensing
galaxies which have been morphologically classified as early-type have
blue mass-to-light ratios which are greater than $5~h$ (Keeton et al.\
1998).  In order for the early-type lensing galaxy in B2319+051 to be
consistent with the measurements from other lenses, we predict that
the source redshift $z_s$ will be less than 1.5.

\section{Conclusion}

As part of a continuing observational program to study gravitational
lens systems, we have measured previously unidentified redshifts in
three lens systems, SBS 0909+532, HST 1411+5211, and CLASS B2319+051.
The spectral characteristics of the central lensing galaxy in all
three systems suggest that each is an early-type galaxy.
High-angular-resolution HST images confirm that these lenses appear as
morphologically normal early-type galaxies (Fischer et al.\ 1998;
Marlow et al.\ 1999; Leh\'ar et al.\ 1999).  The observations suggest,
as previously noted, that the majority of lensing galaxies are
early-types (see Keeton et al.\ 1998 and references therein).  For the
lensing galaxy in HST 1411+5211, we measure a blue mass-to-light ratio
which is a factor of $\sim 5$ larger than the average lensing galaxy
at a similar redshift. The presence of the massive cluster CL 3C295 is
responsible for this significantly enhanced ratio.  

For the other two systems, we are only able to constrain the
mass-to-light ratios. The large observational uncertainties on the
luminosity of the lensing galaxy in SBS 0909+532 allow a wide range in
mass-to-light ratio; however, our measurement is consistent with the
observed values in other high-redshift gravitational lenses.
Similarly for the primary lensing galaxy in B2319+051, we predict a
mass-to-light ratio which is typical of previous lens measurements.
Our imaging indicates that both lenses have a few companion galaxies
within $200~h^{-1}$ kpc which have magnitudes and/or colors typical of
an early-type galaxy at the lens redshift.  Consequently, the primary
lensing galaxy may be associated with a group of galaxies as
previously observed in the lens systems MG 0751+2716, PG 1115+080, and
B1422+231 (Kundi\'c et al.\ 1997b,c; Tonry 1998; Tonry \& Kochanek
1999). We are currently pursuing the group hypothesis for both SBS
0909+532 and B2319+051.

Finally, the expected time delays in all three lens systems are
approximately $100~h^{-1}$ days or less (Oscoz et al.\ 1997; Fischer
et al.\ 1998; Marlow et al.\ 1999), and at least one source
(B2319+051) shows evidence of variability (Marlow et al.\ 1999).
Therefore, some of these systems may be suitable for measuring
$H_{0}$.

\acknowledgements

We would like to thank the referee Emilio Falco for very useful
comments on the text.  We also thank Mark Metzger, Gordon Squires, and
Chuck Steidel for helpful discussions and essential material aids to
this paper. The W.M. Keck Observatory is operated as a scientific
partnership between the California Institute of Technology, the
University of California, and the National Aeronautics and Space
Administration.  It was made possible by generous financial support of
the W. M. Keck Foundation.  The National Radio Astronomy Observatory
is operated by Associated Universities, Inc., under cooperative
agreement with the National Science Foundation.  MERLIN is operated as
a National Facility by NRAL, University of Manchester, on behalf of
the UK Particle Physics and Astronomy Research Council.  Support for
LML was provided by NASA through Hubble Fellowship grant
HF-01095.01-97A awarded by the Space Telescope Science Institute,
which is operated by the Association of Universities for Research in
Astronomy, Inc., for NASA under contract NAS 5-26555. This work was
partially supported by the NSF under grant \#AST 9420018.

\include{tables}

\include{figures}

\end{document}

%% file: tables.tex
\newpage

% Table 1 : Observations

\begin{deluxetable}{rlccrc}
\tablewidth{0pt}
\tablecaption{The Observations}
\tablehead{
\colhead{System} &
\colhead{Date} &
\colhead{$t_{exp}$}  &
\colhead{Slit Width}         &
\colhead{P.A.}	    &
\colhead{Coverage}\\
\colhead{} &
\colhead{} &
\colhead{(sec)}&
\colhead{(arcsec)}&
\colhead{(deg)}&
\colhead{\AA}}
\startdata
SBS 0909+532 	& 1997 Dec 27	& 1200	& 1.0	& 115.2	& 3802 -- 8783 \\
HST 1411+5211	& 1998 Jun 29	& 5400	& 0.7	& 100.6 & 3855 -- 8836 \\
B2319+051 G1	& 1998 Aug 01	& 3600	& 1.0	& 0.0 	& 4009 -- 8991 \\
	G2 	& 1999 Jul 15   & 1500  & 0.7	& 45.0  & 4131 -- 9132 \\
\enddata
\end{deluxetable}

% Table 2 : Detected Spectral Lines

\begin{deluxetable}{lclccccc}
\tablecolumns{6}
\tablewidth{0pt}
\tablecaption{Detected Spectral Lines}
\tablehead{
\colhead{} &
\colhead{} &
\multicolumn{6}{c}{Observed Wavelength (\AA)} \\
\colhead{} &
\colhead{${\rm \lambda_{o}}$} &
\colhead{} &
\colhead{} &
\colhead{} &
\colhead{} &
\colhead{} &
\colhead{} \\
\cline{3-8} \\
\colhead{Ion}   &
\colhead{(\AA)} &
\multicolumn{2}{c}{SBS 0909+532} &
\multicolumn{2}{c}{HST 1411+5211} &
\multicolumn{2}{c}{CLASS 2319+051} \\
\colhead{} &
\colhead{} &
\colhead{Source} &
\colhead{Lens} &
\colhead{Source} &
\colhead{Lens} &
\colhead{Lens 1} &
\colhead{Lens 2}}
\startdata
Ly$\alpha$	& 1216	& \nodata	& \nodata	& 4634		& \nodata	& \nodata	& \nodata \\
\ion{He}{2} 	& 1641 	& 3901		& \nodata	& \nodata 	& \nodata 	& \nodata 	& \nodata \\
\ion{O}{3}]	& 1663	& 3953		& \nodata	& \nodata 	& \nodata 	& \nodata 	& \nodata \\
\ion{C}{3}] 	& 1909	& 4538		& \nodata	& \nodata 	& \nodata 	& \nodata 	& \nodata \\
\ion{C}{2}]	& 2326	& 5529		& \nodata	& \nodata 	& \nodata 	& \nodata 	& \nodata \\
\ion{Fe}{2}	& 2382	& \nodata	& 4359		& \nodata	& \nodata 	& \nodata 	& \nodata \\
		& 2586	& \nodata	& 4732		& \nodata	& \nodata 	& \nodata 	& \nodata \\
		& 2599	& \nodata	& 4756		& \nodata 	& \nodata 	& \nodata 	& \nodata \\
\ion{Mg}{2}	& 2796	& 6646		& 5117		& \nodata 	& \nodata 	& \nodata 	& \nodata \\
\ion{Mg}{1}	& 2853	& \nodata	& 5221		& \nodata	& \nodata 	& \nodata 	& \nodata \\
\ion{O}{3}	& 3133	& 7447		& \nodata	& \nodata 	& \nodata 	& \nodata 	& \nodata \\
\ion{Ne}{5}]	& 3346	& 7953		& \nodata	& \nodata	& \nodata 	& \nodata 	& \nodata \\
		& 3427	& 8146		& \nodata	& \nodata 	& \nodata 	& \nodata 	& \nodata \\
\ion{O}{2}	& 3727	& \nodata 	& \nodata 	& \nodata	& \nodata	& 6053 		& 5921 \\
\ion{Ca}{2} K	& 3934	& \nodata	& 7199		& \nodata 	& 5758		& 6388		& 6250 \\
\ion{Ca}{2} H	& 3968	& \nodata	& 7261		& \nodata	& 5814		& 6422		& 6308 \\
H$\delta$	& 4102 	& \nodata	& \nodata	& \nodata	& 6013		& 6661		& 6511 \\
H$\gamma$	& 4341	& \nodata	& \nodata	& \nodata	& \nodata	& 7046		& \nodata \\
H$\beta$	& 4841	& \nodata	& \nodata	& \nodata	& 7125		& \nodata	& 7729 \\
	
\enddata
\end{deluxetable}

% Table 3 : Lens System Parameters

\begin{deluxetable}{rccccccc}
\scriptsize
\tablewidth{0pt}
\tablecaption{Lens System Parameters}
\tablehead{
\colhead{System} 	&
\colhead{$z_{\ell}$} 	&
\colhead{$z_s$}  	&
\colhead{$D_l$}         &
\colhead{$D_s$}	    	&
\colhead{$D_{ls}$}	&
\colhead{$M_E$}		&
\colhead{${(M/L)}_{B}$}	\\
\colhead{} &
\colhead{} &
\colhead{} &
\colhead{($h^{-1}$ Mpc)}&
\colhead{($h^{-1}$ Mpc)}&
\colhead{($h^{-1}$ Mpc)}&
\colhead{($10^{11}~h^{-1}~M_{\odot}$)}&
\colhead{($h~{[M/L]}_{\odot}$)}}
\startdata
SBS 0909+532 	& $0.8302 \pm 0.0001$ & $1.3764 \pm 0.0003$ & $999 \pm 0.02$ & $1129 \pm 0.01$ & $301 \pm 0.09$ & $1.42 \pm 0.03$ & $4^{+11}_{-3}$\\
HST 1411+5211	& $0.4641 \pm 0.0001$ & $2.811 \pm 0.005$ & $778 \pm 0.07$ & $1160 \pm 0.48$ & $730 \pm 0.13$ & $1.98 \pm 0.02$ & $41.3 \pm 1.2$\\
B2319+051	& $0.6238 \pm 0.0001$ & \nodata & $896 \pm 0.04$ & \nodata & \nodata & \nodata & \nodata \\
\enddata
\end{deluxetable}

%% file: figures.tex
\def\plottwo#1#2{\centering \leavevmode
    \epsfxsize=.45\columnwidth \epsfbox{#1} \hfil
    \epsfxsize=.45\columnwidth \epsfbox{#2}}

\newpage

\begin{figure}
\centerline{\epsfbox{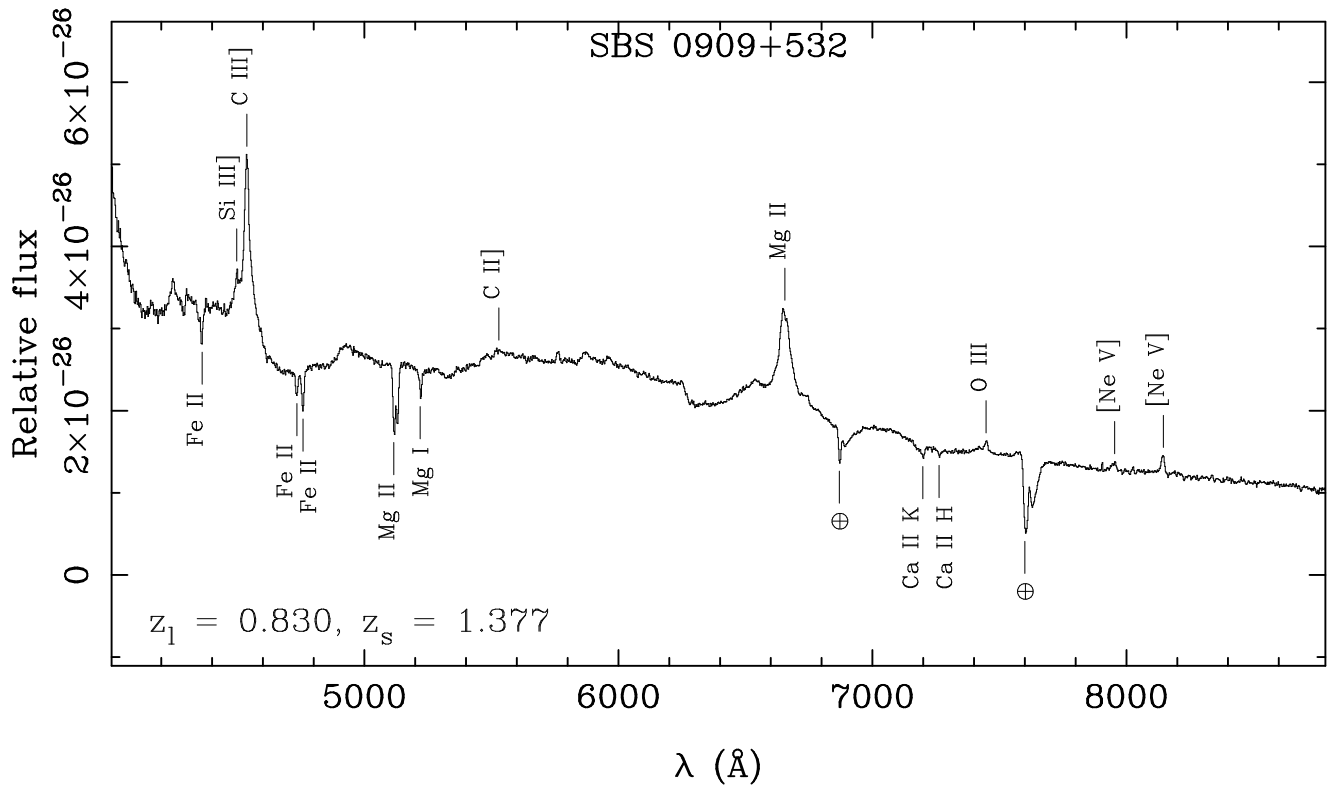}}
\caption{LRIS spectrum of the SBS 0909+532 system.  Flux 
calibration using the spectrophotometric standard Feige 34 has been
performed.  The vertical axis has been converted from $F_{\lambda}$ to
$F_{\nu}$ in order to emphasize the absorption features associated
with the lensing galaxy.  Spectral lines from both the lensing galaxy
($z_{\ell} = 0.830$) and the background source ($z_s = 1.377$) are
seen in the spectrum.  All marked non-terrestrial absorption lines are
due to the lensing galaxy, while all marked emission lines are due to
the background source.  The previously unmeasured lensing redshift is
determined from the \ion{Ca}{2} H \& K features.}
\label{0909spec}
\end{figure}

\begin{figure}
\centerline{\epsfbox{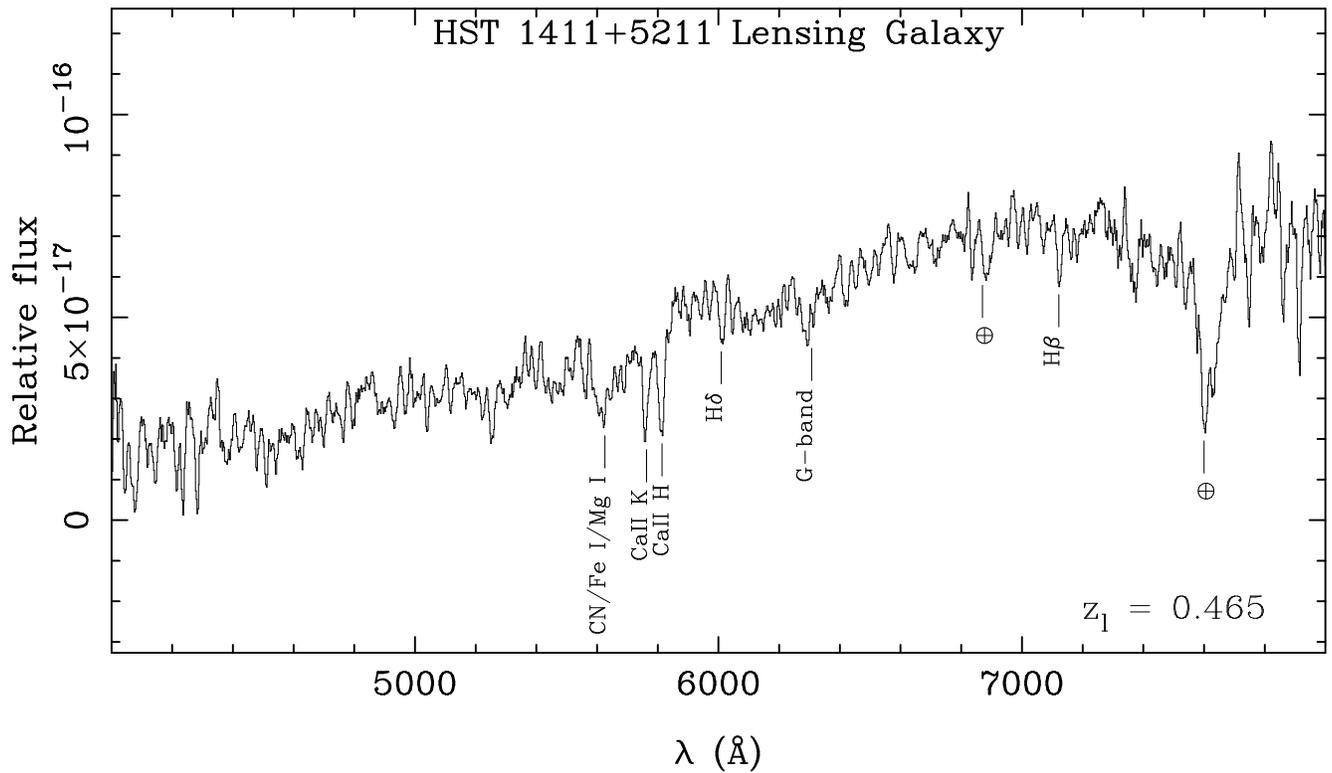}}
\caption{LRIS spectrum of the lensing galaxy in HST 1411+5211.  Flux 
calibration using the spectrophotometric standard G138-31 has been
performed.  The spectrum has been smoothed with a box car of size
12\AA. The previously unmeasured lensing redshift of $z_{\ell}=0.465$
is determined from the \ion{Ca}{2} H \& K, H$\delta$, and H$\beta$
features.}
\label{1411lens}
\end{figure}

\begin{figure}
\centerline{\epsfbox{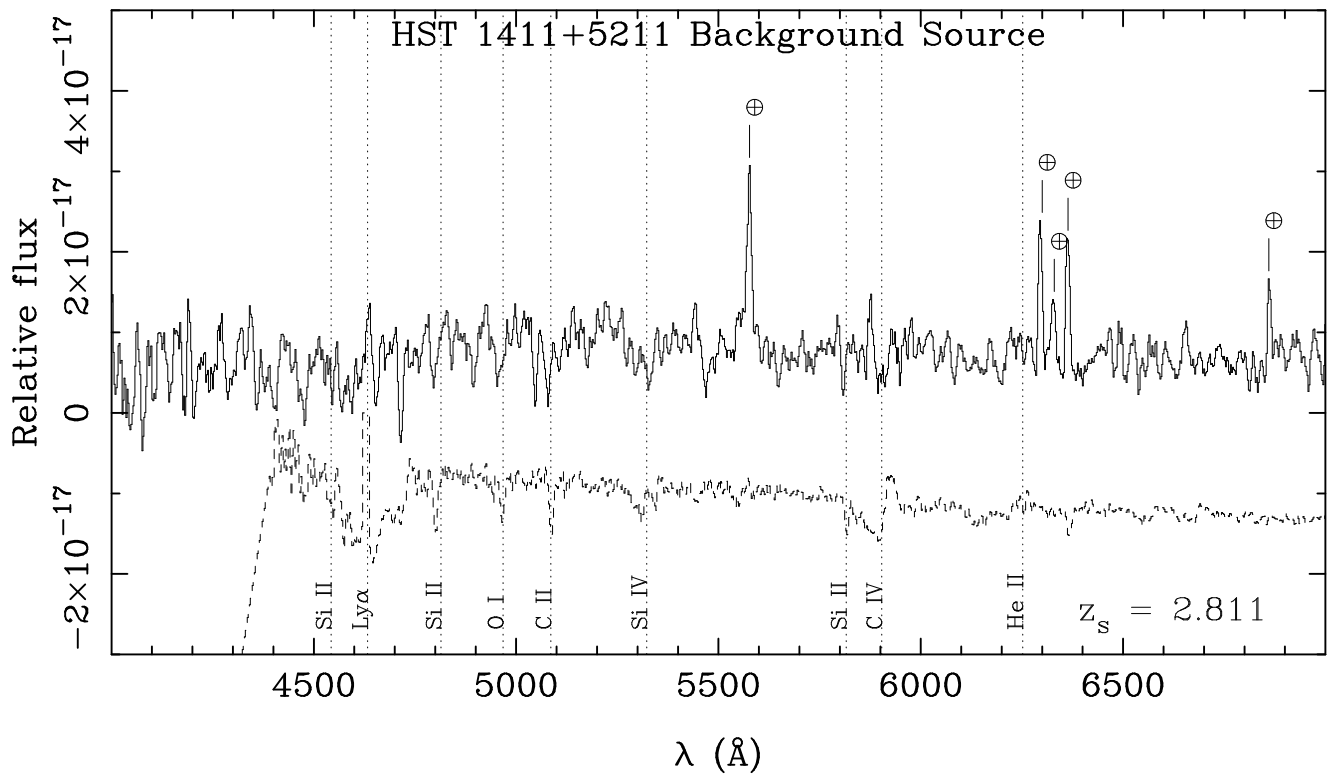}}
\caption{LRIS spectrum of the background source in HST 1411+5211.  Flux 
calibration using the spectrophotometric standard G138-31 has been
performed. The spectrum has been smoothed with a box car of size
12\AA. The previously unmeasured source redshift of $z_s=2.811$ is
determined from the Ly$\alpha$ line and several observed absorption
features.  Below this spectrum we plot a scaled spectrum of the nearby
starburst galaxy NGC 4214 (Leitherer et al.\ 1996).  The position of
several stellar and interstellar features which are routinely observed
in both nearby and distant star-forming galaxies are indicated with
vertical lines.}
\label{1411source}
\end{figure}

\begin{figure}
\centerline{\epsfbox{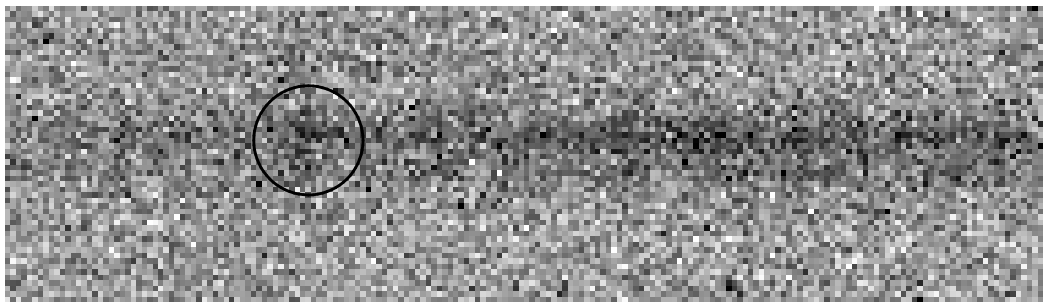}}
\caption{The two-dimensional, sky-subtracted spectrum of HST 1411+5211.  The
brighter, upper trace belongs to the lensing galaxy.  The fainter,
lower trace belongs to the source. The dispersion axis ranges from
4500\AA\ to 4965\AA, and the spatial axis covers 11\farcs{8}.
Emission from the Ly$\alpha$ line at the observed wavelength of
4643\AA\ is circled. Note that the Ly$\alpha$ emission appears
significantly extended.}
\label{1411s2D}
\end{figure}

\begin{figure}
\epsfysize=4in
\centerline{\epsfbox{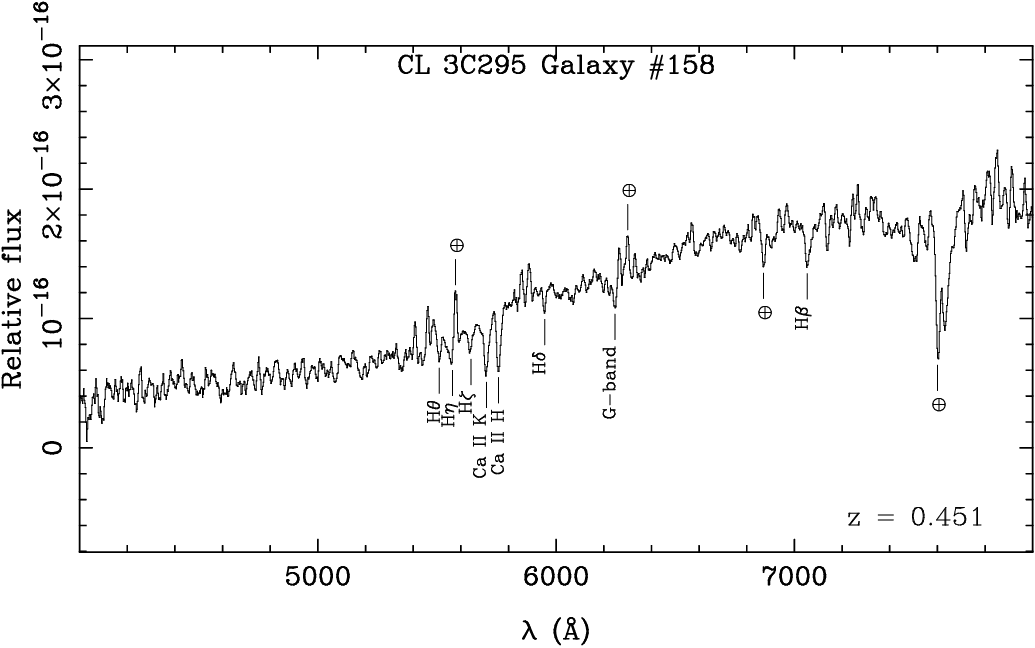}}
\vskip 1.5cm
\epsfysize=4in
\centerline{\epsfbox{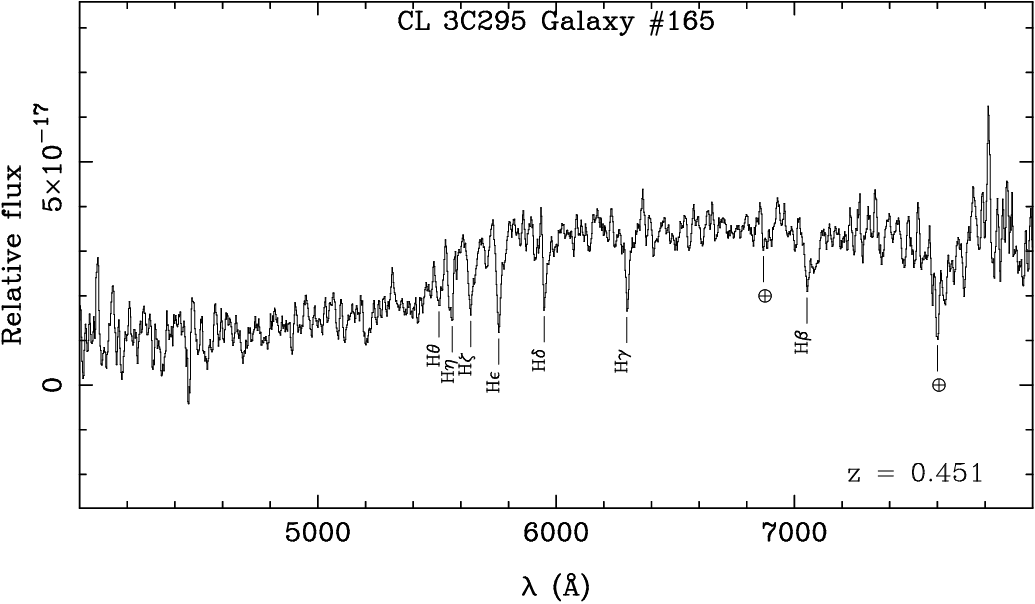}}
\caption{LRIS spectra of two cluster members in CL 3C295 (CL1409+5226).
Galaxy \#158 ({\it upper panel}) and galaxy \#165 ({\it lower panel})
of Table 6 in Dressler \& Gunn (1992). Flux calibration using the
spectrophotometric standard G138-31 has been performed.  Both spectra
have been smoothed with a box car of size 12\AA. The galaxy
redshifts are determined from several spectral lines which include
\ion{Ca}{2} H \& K, H$\theta$, H$\eta$, H$\zeta$, H$\epsilon$, H$\delta$, H$\gamma$, 
and H$\beta$.}
\label{1411gal}
\end{figure}

\begin{figure}
\epsfysize=3.8in
\centerline{\epsfbox{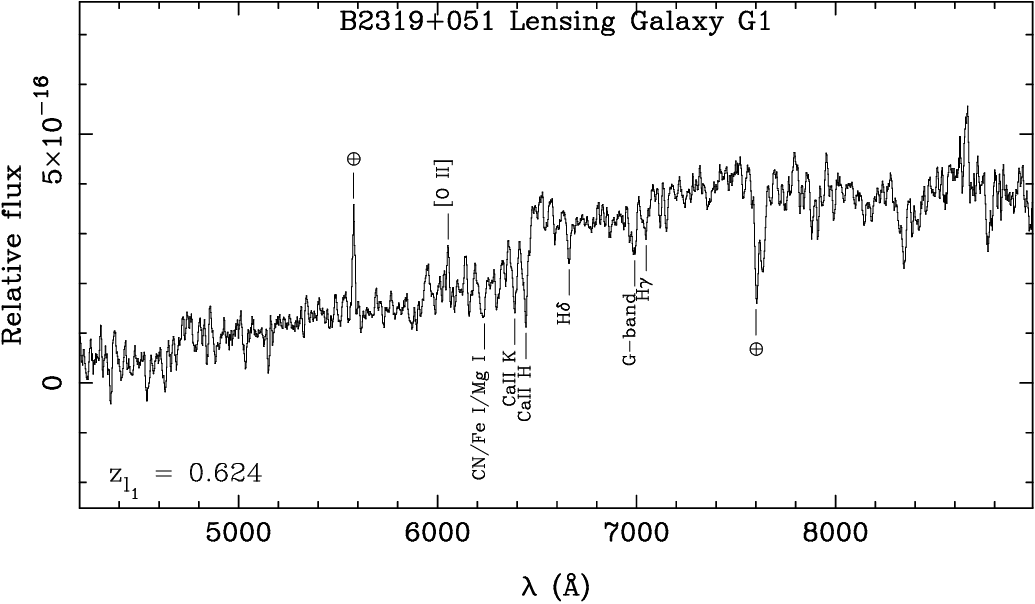}}
\vskip 1.5cm
\epsfysize=3.8in
\centerline{\epsfbox{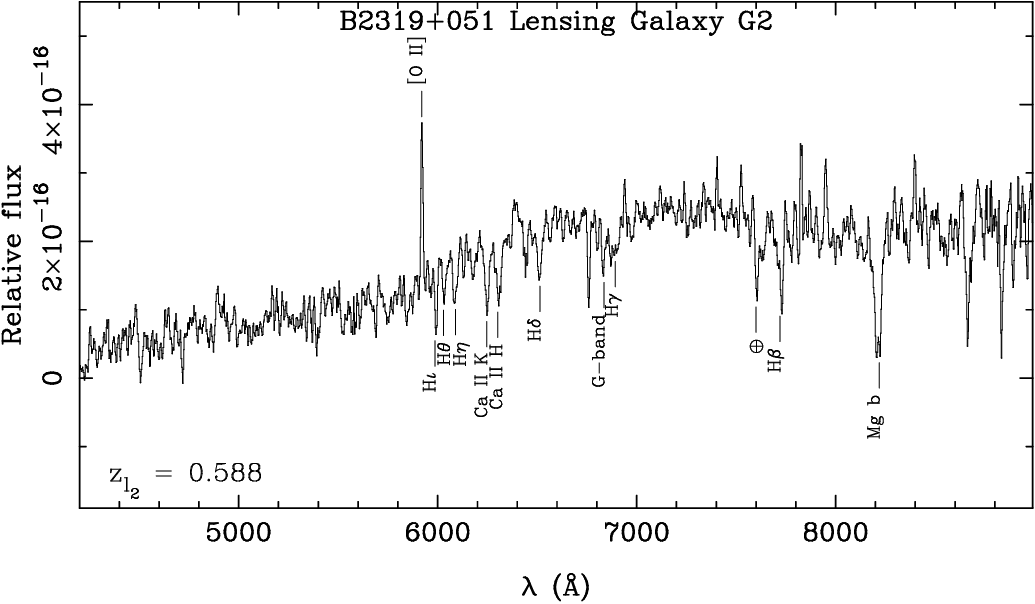}}
\caption{LRIS spectra of the lensing galaxies G1 ({\it upper panel}) and
G2 ({\it lower panel}) in the CLASS B2319+051 system. Flux calibration
using the spectrophotometric standards BD332641 for G1 and Feige 110
for G2 has been performed. The spectrum has been smoothed with a box
car the size of 12\AA. The previously unmeasured redshifts of
$z_{{\ell}_{1}} = 0.624$ and $z_{{\ell}_{2}} = 0.588$ are determined
from several spectral lines which include [\ion{O}{2}], \ion{Ca}{2} H
\& K, H$\delta$, and H$\gamma$.}
\label{2319spec} 
\end{figure}

\begin{figure}
\centerline{\epsfbox{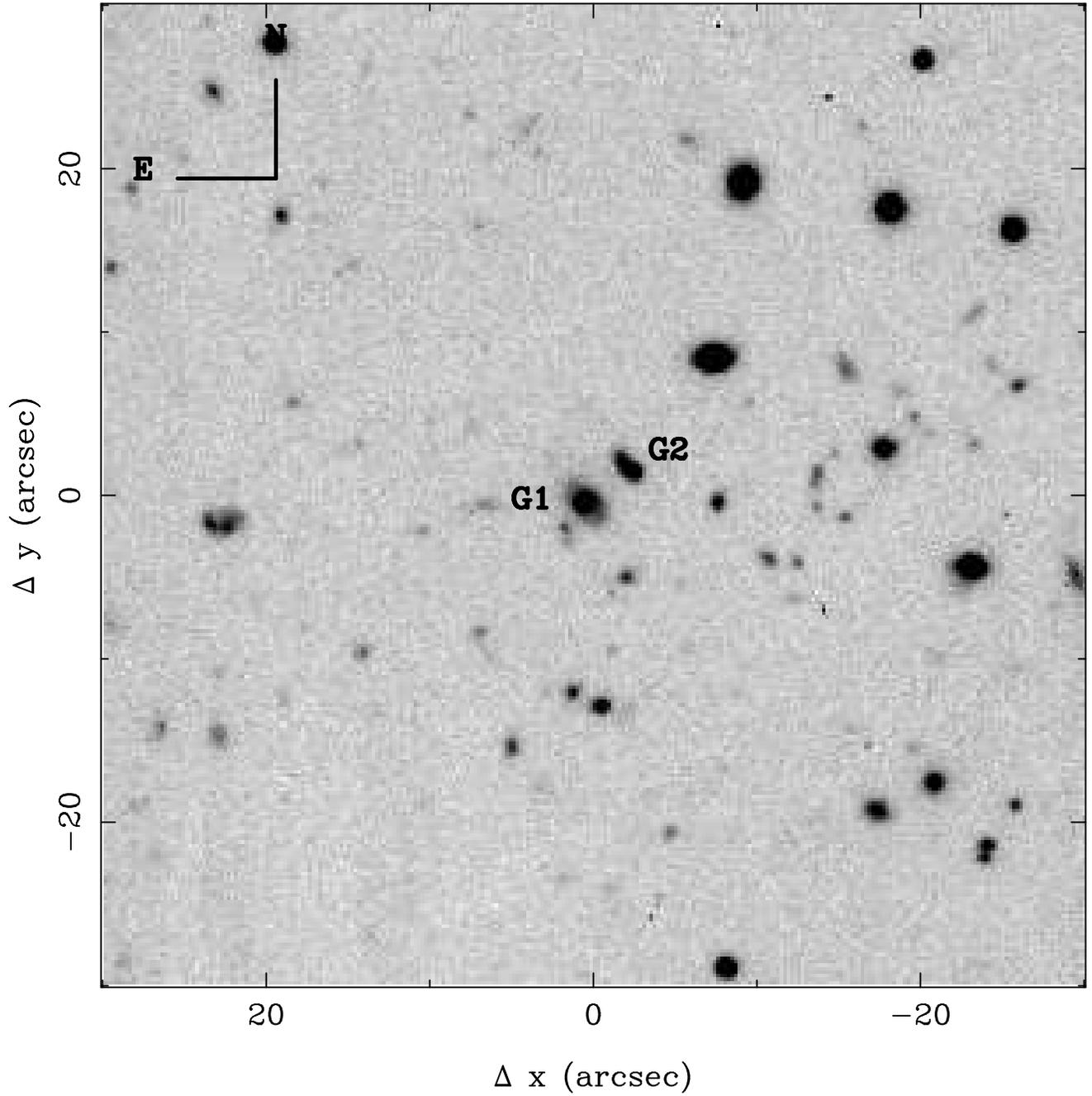}}
\caption{The composite $R$ band image centered on the gravitational lens
B2319+052.  The field-of-view is $1' \times 1'$, and the total
exposure time is 1200 sec.  The lensing galaxies G1 and G2 are
labeled.}
\label{2319im}
\end{figure}